\documentclass[letterpaper,twocolumn,10pt]{article}
\usepackage{usenix2019,epsfig,endnotes}
\usepackage{amsmath}
\usepackage{amsfonts}
\usepackage{todonotes}
\newcommand\blfootnote[1]{%
  \begingroup
  \renewcommand\thefootnote{}\footnote{#1}%
  \addtocounter{footnote}{-1}%
  \endgroup
}

\usepackage{appendix}
\usepackage{graphicx}
\usepackage{subfig}
\usepackage{chngcntr}
\usepackage{hyperref}

\begin{document}

%\listoftodos{}

%don't want date printed
\date{}

%make title bold and 14 pt font (Latex default is non-bold, 16 pt)
\title{\Large \bf ALOHA: Auxiliary Loss Optimization for Hypothesis Augmentation\vspace{-1.5em}
}

%for single author (just remove % characters)
\author{
{\rm Ethan M. Rudd$^{*}$, Felipe N. Ducau$^{*}$, Cody Wild, Konstantin Berlin, and Richard Harang$^{*}$}\\
Sophos PLC
%{\rm Authors Anonymized for Review}\\
%nstitution Anonymized for Review
%\and
%{\rm Second Name}\\
%Second Institution
% copy the following lines to add more authors
% \and
% {\rm Name}\\
%Name Institution
\vspace{-2em}
} % end author

\maketitle

% Use the following at camera-ready time to suppress page numbers.
% Comment it out when you first submit the paper for review.
\thispagestyle{empty}

\section*{Abstract}

Malware detection is a popular application of Machine Learning for Information Security (ML-Sec), in which an ML classifier is trained to predict whether a given file is malware or benignware. Parameters of this classifier are typically optimized such that outputs from the model over a set of input samples most closely match the samples’ true malicious/benign (1/0) target labels. However, there are often a number of other sources of contextual metadata for each malware sample, beyond an aggregate malicious/benign label, including multiple labeling sources and malware type information (e.g., ransomware, trojan, etc.),  which we can feed to the classifier as auxiliary prediction targets. In this work, we fit deep neural networks to multiple additional targets derived from metadata in a threat intelligence feed for Portable Executable (PE) malware and benignware, including a multi-source malicious/benign loss, a count loss on multi-source detections, and a semantic malware attribute tag loss. We find that incorporating multiple auxiliary loss terms yields a marked improvement in performance on the main detection task. We also demonstrate that these gains likely stem from a more informed neural network representation and are not due to a regularization artifact of multi-target learning. Our auxiliary loss architecture yields a significant reduction in detection error rate (false negatives) of 42.6\% at a false positive rate (FPR) of $10^{-3}$ when compared to a similar model with only one target, and a decrease of 53.8\% at $10^{-5}$ FPR.

\vspace{-1em}

%old abstract
%In applications of Machine Learning for Information Security (ML-Sec), malware detection is commonly approached by training a classifier with a single binary malicious/benign target.
%This is true even when far more label information is typically already available, including multiple label sources or statistics on those multiple sources, e.g., number of detections across all sources. 
%In this work, we revisit multi-target learning in a deep neural network for the task of malware detection by fitting a multi-objective loss function to multiple additional targets derived from metadata in a threat intelligence feed. 
%We experiment with static analysis of Portable Executable (PE) files by adding a multi-source malicious/benign loss, a count loss on multi-source detections, and a semantic malware attribute tag loss. We find that incorporating multiple auxiliary loss terms yields a marked improvement in performance on the main task. We also demonstrate that these gains likely stem from a more informed neural network representation and are not due to a regularization artifact of multi-objective learning. Our auxiliary loss architecture yields a significant
%reduction in detection error rate (false negatives) of 42.6\% at a false positive rate (FPR) of $10^{-3}$ when compared to a similar model with only one target, and a decrease of 53.8\% at $10^{-5}$ FPR. 

%%%%%%%%%%%%%%%%%%%%%%%%%%%%%%%%%%%%%%%%%%%%%%
% 1. INTRODUCTION
%%%%%%%%%%%%%%%%%%%%%%%%%%%%%%%%%%%%%%%%%%%%%%
\section{Introduction}

Machine learning (ML) for computer security (ML-Sec) has proven to be a powerful tool for malware detection\blfootnote{$*$ Equal contribution.}.\blfootnote{Contact: \url{<first name>.<last name>@sophos.com}} ML models are now integral parts of  commercial anti-malware engines and multiple vendors in the industry have dedicated ML-Sec teams. For the malware detection problem, these models are typically tuned to predict a binary label (malicious or benign) using features extracted from sample files. Unlike signature engines, where the aim is to reactively blacklist/whitelist malicious/benign samples that hard-match  manually-defined  patterns (signatures), ML engines employ numerical optimization on parameters of highly parametric models that aim to learn more general concepts of \textit{malware} and \textit{benignware}. This allows some degree of proactive detection of previously unseen malware samples that is not typically provided by signature-only engines.

Frequently, malware classification is framed as a binary classification task using a simple binary cross-entropy or two-class softmax loss function. 
However, there often exist substantial metadata available at training time that contain more information about each input sample than just an aggregate label of whether it is malicious or benign.  
Such metadata might include malicious/benign labels from multiple sources (e.g., from various security vendors), malware family information, file attributes, temporal information, geographical location information, counts of affected endpoints, and associated tags.  In many cases this metadata will not be available when the model is deployed, and so in general it is difficult to include this data as features in the model (although see Vapnik et al. \cite{vapnik2015learning, vapnik2009new} for one approach to doing so with Support Vector Machines).

It is a popular practice in the domain of malware analysis to derive binary malicious/benign labels based on a heuristic combination of multiple detection sources for a given file, and then use these noisy labels for training ML models \cite{2018arXiv181007260D}. However, there is nothing that precludes training a classifier to predict each of these source labels simultaneously optimization classifier parameters over these predictions + labels. In fact, one might argue intuitively that guiding a model to develop representations capable of predicting multiple targets  simultaneously may have a smoothing or regularizing effect conducive to generalization, particularly if the auxiliary targets are related to the main target of interest. These auxiliary targets can be ignored during test time if they are ancillary to the task at hand (and in many cases the extra weights required to produce them can be removed from the model prior to deployment), but nevertheless, there is much reason to believe that forcing the model to fit multiple targets simultaneously can improve performance on the key output of interest. In this work, we take advantage of multi-target learning \cite{abu1990learning} by exploring the use of metadata from threat intelligence feeds as auxiliary targets for model training.

Research in other domains of applied machine learning supports this intuition \cite{jaderberg2014deep, ranjan2017hyperface, huang2015unconstrained, wu2015weakly, abdulnabi2015multi, rudd2016moon}, however outside of the work of Huang et al. \cite{huang2016mtnet}, multi-target learning has not been widely applied in anti-malware literature. In this paper, we present a wide-ranging study applying auxiliary loss functions to anti-malware classifiers. In contrast to \cite{huang2016mtnet}, which studies the addition of a single auxiliary label for a fundamentally different task, i.e., malware family classification -- we study both the addition of multiple label sources for the same task and multiple label sources for multiple separate tasks. Also, in contrast to \cite{huang2016mtnet}, we do not presume the presence of all labels from all sources, and introduce a per-sample weighting scheme on each loss term to accommodate missing labels in the metadata. We further explore the use of multi-objective training as a way to expand the number of trainable samples in cases where the aggregate malicious/benign label is unclear, and where those samples would otherwise be excluded from purely binary training.  

Having established for which loss types and in which contexts auxiliary loss optimization works well, we then explore \textit{why it works well}, via experiments designed to test whether performance gains are a result of a regularization effect from multi-objective training or information from the content of the target labels that the network has learned to correlate.

In summary, this paper makes the following contributions:
\begin{itemize}

\item A demonstration that including  auxiliary losses yields improved performance on the main task. When all of our auxiliary loss terms are included, we see a reduction of $53.8\%$ in detection error (false negative) rate at $10^{-5}$ false positive rate (FPR) and a $42.6\%$ reduction in detection error rate at $10^{-3}$ FPR compared to our baseline model. We also see a consistently better and lower-variance ROC curve across all false positive rates.

\item A breakdown of performance improvements from different auxiliary loss types. We find that an auxiliary Poisson loss on detection counts tends to yield improved detection rates at higher FPR areas ($\geq 10^{-3}$) of the ROC curve, while multiple binary auxiliary losses tend to yield improved detection performance in lower FPR areas of the ROC curve ($<10^{-3}$). When combined we see a net improvement across the entire ROC curve over using any single auxiliary loss type.

\item An investigation into the mechanism by which multi-objective optimization yields enhanced performance, including experiments to assess possible regularization effects.

% \item Guidance to the ML-Sec community about how auxiliary loss functions may be potentially employed for other problems.
\end{itemize}

The remainder of this paper is laid out as follows: First, in Section \ref{sec:mlpipeline} we discuss some of the metadata available for use as auxiliary targets, and feature extraction methods for portable executable (PE) files.  We then provide details on how we converted the available metadata into auxiliary targets and losses, as well as how the individual losses are weighted and combined on a per-sample basis in Section \ref{sec:implementation}.  We finish that Section with a description of how our dataset was collected and provide summary statistics.  In Section \ref{sec:experiments} we describe our experimental evaluations across a number of combinations of auxiliary targets, and demonstrate the impact of fitting these targets on the performance of the main malware detection task.  Section \ref{sec:discussion} presents discussion of our results, as well as another set of experiments on synthetic targets to explore potential explanations for the observed improvement.  Section \ref{sec:relatedwork} presents related work and Section \ref{sec:conclusion} concludes.

\vspace{-1em}
\section{ML-Sec Detection Pipelines: From Single Objective to Multi-Objective}
\vspace{-1em}
\label{sec:mlpipeline}
In the following, we describe a simplified ML-Sec pipeline for training a malicious file classifier, and propose a simple extension that allows the use of metadata available during training (but not at test time) and improves performance on the classification task.

ML-Sec detection pipelines use powerful machine learning models that require many labeled malicious and benign samples to train. Data is typically gathered from several sources, including deployed anti-malware products and vendor aggregation services, which run uploaded samples through vendor products and provide reports containing per-vendor detections and metadata. The exact nature of the metadata varies, but typically, malicious and benign scores are provided for each of $M$ individual samples on a per-vendor basis, such that, given $V$ vendors, between 0 and $V$ of them will designate a sample malicious. For a given sample, some vendors may choose not to answer, resulting in a missing label for that vendor. Furthermore, many vendors also provide a detection name (or malware family name) when they issue a detection for a given file. Additional information may also be available, but crucially, the following metadata are presumed present for the models presented in this paper: i) per-vendor labels for each sample, either malicious/benign (mapped to binary $1/0$, respectively) or NULL; ii) textual per-vendor labels on the sample describing the family and variant of the malware (an empty string if benign); and iii) time at which the sample was first seen.

Using the individual vendor detections, an aggregate label can be derived either by a voting mechanism or by thresholding the net number of vendors that identify a given sample as malicious. The use of aggregated anti-malware vendor detections as a noisy labeling source presumes that the vendor diagnoses are  generally accurate. While this is not necessarily a valid assumption, e.g., for novel malware and benignware, this is typically accounted for by using samples and metadata that are slightly dated so that vendors can correct their respective mistakes (e.g., by blacklisting samples in a signature database). 

Each malware/benignware sample must also be converted to a numerical vector to be able to train a classifier, a process called \textit{feature extraction}. In this work we focus on static malware detection, meaning that we assume only access to the binary file, as opposed to dynamic detection, in which the features used predominantly come from the execution of the file. The feature extraction mechanism varies depending on the format type at hand, but consists of some numerical transformation that preserves aggregate and fine-grained information throughout each sample, for example, the feature extraction proposed by Saxe et al. \cite{saxe2015deep} -- which we use in this work -- uses windowed byte statistics, 2D histograms of delimited string hash vs. length, and histograms of hashes of PE-format specific metadata -- e.g., imports from the import address table (IAT). 

Given extracted features and derived labels, a classifier is then trained. Parameters are tuned to minimize mis-classification as measured by some loss criterion, which under the constraints of some statistical noise model measures the deviation in predictions from their ground truth. For both neural networks and ensemble methods a logistic sigmoid is commonly used to constrain predictions to a [0,1] range, and a cross-entropy loss between predictions and labels is used as the minimization criterion under a Bernoulli noise model per-label.

\begin{figure}[!t]
    \centering
    \includegraphics[width=\linewidth]{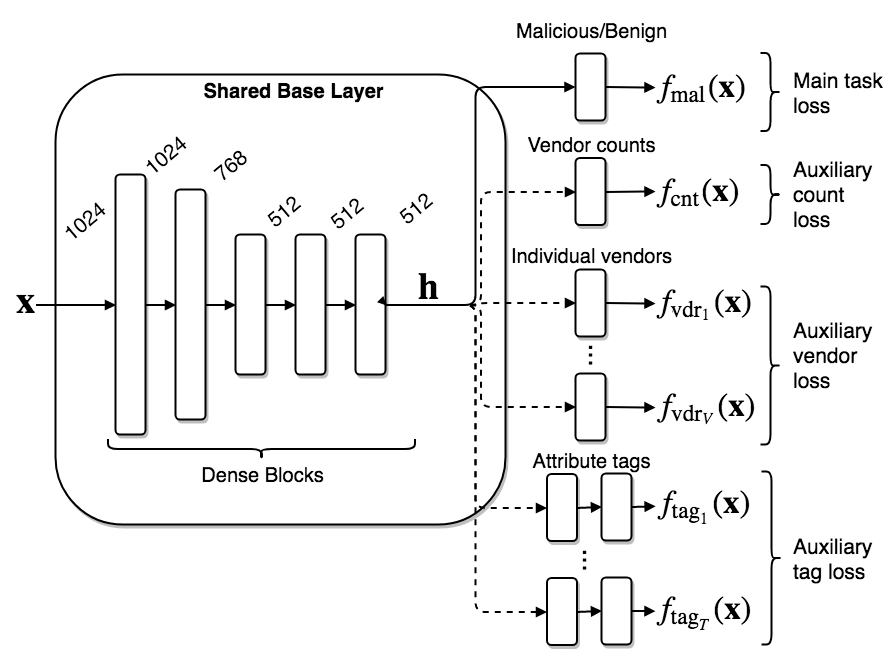}
    \caption{A schematic overview of our neural network architecture. Multiple output layers with corresponding loss functions are optionally connected to a common base topology which consists of five dense blocks. Each block is composed of a Dropout, dense and batch normalization layers followed by an exponential linear unit (ELU) activation of sizes 1024, 768, 512, 512, and 512. This base,  connected to our main malicious/benign output (solid line in the figure) with a loss on the aggregate label constitutes our baseline architecture. Auxiliary outputs and their respective losses are represented in dashed lines. %In this paper we explore the effects on its performance of adding auxiliary outputs/losses (optional dashed lines).
    The auxiliary losses fall into three types: count loss, multi-label vendor loss, and multi-label attribute tag loss. The formulation of each of these auxiliary loss types is explained in Section \ref{sec:implementation}.\vspace{-2em}}
    \label{fig:schematic}
\end{figure}

While the prior description roughly characterizes ML-Sec pipelines discussed in literature to date, note that much information in the metadata, which is often not used to determine the sample label but \emph{is} correlated to the aggregate classification, is not used in training, e.g., the individual vendor classifications, the combined number of detections across all vendors, and information related to malware type that could be derived from the detection names. In this work, we augment a deep neural network malicious/benign classifier with additional output predictions of vendor counts, individual vendor labels, and/or attribute tags. These separate prediction arms were given their own loss functions which we jointly minimized through backpropagation. The difference between a conventional malware detection pipeline and our model can be visualized by considering Figure \ref{fig:schematic} in the absence and presence of auxiliary outputs (and their associated losses) connected by the dashed lines. In the next section, we shall explore the precise formulation and implementation of these different loss functions.

%%%%%%%%%%%%%%%%%%%%%%%%%%%%%%%%%%%%%%%%%%%%%%
% 3. IMPLEMENTATION DETAILS
%%%%%%%%%%%%%%%%%%%%%%%%%%%%%%%%%%%%%%%%%%%%%%
\vspace{-1em}
\section{Implementation Details}
\vspace{-1em}
\label{sec:implementation}
In this section we describe our implementation of the experiments sketched above. We first introduce our model immediately below, followed by the various loss functions -- denoted by $L_{\text{loss type}}\left(X,Y\right)$ for some input features $X$ and targets $Y$ -- associated with the various outputs of the model, as well as how the labels $Y$ representing the targets of these outputs are constructed. Finally we introduce how our data set of $M$ samples associated with $V$ vendor targets is collected. We use the same feature representation as well as the same general model class and topology for all experiments. Each portable executable file is converted into a feature vector as described in \cite{saxe2015deep}.

The base for our model (shown in Figure \ref{fig:schematic}) is a feedforward neural network incorporating multiple blocks composed of Dropout \cite{srivastava2014dropout}, a dense layer, batch normalization \cite{ioffe2015batch}, and an exponential linear unit (ELU) activation \cite{clevert2015fast}.  The core of the model contains five such blocks with 1024, 768, 512, 512, and 512 hidden units, respectively. This base topology applies the function $f(\cdot)$ to the input vector to produce an intermediate 512 dimensional representation of the input file $\mathbf{h} = f(\mathbf{x})$. We then append to this model an additional block, consisting of one or more dense layers and activation functions, for each output of the model.  We denote the composition of our base topology and our target-specific final dense layers and activations applied to features $\mathbf{x}$ by $f_{\text{target}}(\mathbf{x})$. The output for the main malware/benign prediction task -- $f_{\text{mal}}(\mathbf{x})$ -- is always present and consists of a single dense layer followed by a sigmoid activation on top of the base shared network that aims to estimate the probability of the input sample being malicious. A network architecture with only this malware/benign output serves as the baseline for our evaluations. To this baseline model we add auxiliary outputs with similar structure as described above: one fully connected layer (two for the \emph{tag} prediction task in Section \ref{sec:tags_loss}) which produces some task-specific number of outputs (a single output, with the exception of the restricted generalized Poisson distribution output, which uses two) and some task-specific activation described in the associated sections below. 

Except where noted otherwise, all multi-task losses were produced by computing the sum, across all tasks, of the per-task loss multiplied by a task-specific weight (1.0 for the malware/benign task and 0.1 for all other tasks; see Section \ref{sec:experiments}). Training was standardized at 10 epochs; for all experiments we used a training set of 9 million samples and a test set of approximately 7 million samples.  Additional details about the training and test data are reported in Section \ref{sec:dataset}. Additionally, we used a validation set of 100,000 samples to ensure that each network had converged to an approximate minimum on validation loss after 10 epochs. All of our models were implemented in Keras \cite{chollet2015keras} and optimized using the Adam optimizer \cite{kingma2014adam} with Keras's default parameters.

\subsection{Malware Loss}
\label{sec:mal_loss}
As explained in Section \ref{sec:mlpipeline}, for the task of predicting if a given binary file, represented by its features $\mathbf{x}^{(i)}$ is malicious or benign we used a binary cross-entropy loss between the malware/benign output of the network $\hat{y}^{(i)} = f_{\text{mal}}(\mathbf{x}^{(i)})$  and the malicious label $y^{(i)}$. This results in the following loss for a dataset with $M$ samples:

\begin{align}
\label{eq:main_mal_loss}
    L_{\text{mal}}(X, Y) &= \frac{1}{M} \sum_{i=1}^{M} \ell_{\text{mal}}(f_{\text{mal}}(\mathbf{x}^{(i)}), y^{(i)}) \nonumber \\
              &= - \frac{1}{M} \sum_{i=1}^{M} y^{(i)} \log(\hat{y}^{(i)}) + 
                                             (1 - y^{(i)})\log(1 - \hat{y}^{(i)}).
\end{align}

In this paper, we use a ``1-/5+'' criterion for labeling a given file as malicious or benign: if a file has one or fewer vendors reporting it as malicious, we label the file as `benign' and use a weight of 1.0 for the malware loss for that sample.  Similarly, if a sample has five or more vendors reporting it as malicious, we label the file as `malicious' and use a weight of 1.0 for the malware loss for that sample.  %If a sample has two, three, or four positive results, then we take the file as unlabeled and assign a sample weight of 0.0 to the malware loss for that sample.

\subsection{Vendor Count Loss}
%In order to
To more finely distinguish between positive results, we investigate the use of the total number of `malicious' reports for a given sample from the vendor aggregation service as an additional target; the rationale being that a sample with a higher number of malicious vendor reports should, all things being equal, be more likely to be malicious.  In order to properly model this target, we require a suitable noise model for count data. A popular candidate is a Poisson noise model, parameterized by a single parameter $\mu$, which assumes that counts follow a Poisson process, where $\mu$ is the mean and variance of the Poisson distribution. The probability of an observation of $y$ counts conditional on $\mu$ is 

\begin{equation}
\label{eq:poisson_dist}
P(y|\mu) = \mu^y e^{-\mu}/y!.
\end{equation}
\noindent
In our problem, as we expect the mean number of positive results for a given sample to be related to the file itself, we attempt to learn to \emph{estimate} $\mu$ conditional on each sample $\mathbf{x}^{(i)}$ in such a way that the likelihood of $y^{(i)}|\mu^{(i)}$ is maximized (or, equivalently, the negative log-likelihood is minimized).  Denote the output of the neural network with which we are attempting to estimate the mean count of vendor positives for sample $i$ as $f_{\text{cnt}}(\mathbf{x}^{(i)})$. Note that under a non-distributional loss, this would be denoted by $\hat{y}^{(i)}$, however since we are fitting a parameter of a distribution, and not the sample label $y$ directly, we use different notation in this section. By taking some appropriate activation function $a(\cdot)$ that maps $f_{\text{cnt}}(\mathbf{x}^{(i)})$ to the non-negative real numbers, we can write $\mu^{(i)} = a\left(f_{\text{cnt}}(\mathbf{x}^{(i)})\right)$.  Consistent with generalized linear model (GLM) literature \cite{mccullagh2018generalized}, we use an exponential activation for $a$, though one could equally well employ some other transformation with the correct range, for instance the ReLU function. 

Letting $y^{(i)}$ here denote the actual number of vendors that recognized sample $\mathbf{x}^{(i)}$ as malicious, the corresponding negative log-likelihood loss over the dataset is

\begin{align}
\label{eq:poisson_loss}
    L_{\text{p}}(X, Y) &= \frac{1}{M} \sum_{i=1}^M
    \ell_{\text{p}} \left(a \left(f_{\text{cnt}}(\mathbf{x}^{(i)})\right), y^{(i)} \right) \nonumber \\
    &= \frac{1}{M} \sum_{i=1}^M
    \mu^{(i)} - y^{(i)} \log(\mu^{(i)}) + \log(y^{(i)}!),
\end{align}

\noindent
which we will refer to as the \emph{Poisson} or \emph{vendor count} loss.
In practice, we ignore the $\log(y^{(i)}!)$ term when minimizing this loss since it does not depend on the parameters of the network.  

A Poisson loss is more intuitive for dealing with count data than other common loss functions, even for count data not generated by a Poisson process. This is partly due to the discrete nature of the distribution and partly because the assumption of increased variance with predicted mean is more accurate  than a homoscedastic -- i.e., constant variance -- noise model. 

While the assumption of increasing variance with predicted count value seems reasonable, it is very unlikely that vendor counts perfectly follow a Poisson process -- where the mean \textit{is} the variance -- due to correlations between vendors, which might occur from modeling choice and licensing/OEM between vendor products. The variance might increase at a higher or lower rate than the count and might not even be directly proportional to or increase monotonically with  the count. Therefore, we also implemented a Restricted Generalized Poisson distribution \cite{famoye1993restricted} -- a slightly more intricate noise model that accommodates dispersion in the variance of vendor counts. Given dispersion parameter $\alpha$, the Restricted Generalized Poisson distribution has a probability mass function (pmf):

\begin{equation}
\label{eq:gen_poisson_dist}
    P(y|\alpha, \mu) = \left( \frac{\mu}{1+\alpha \mu} \right)^y (1 + \alpha y)^{y-1} \exp\left( \frac{-\mu(1 + \alpha y)}{1 + \alpha \mu} \right)/y!.
\end{equation} 

When $\alpha=0$, this reduces to Eq. \ref{eq:poisson_dist}. $\alpha>0$ accounts for over-dispersion, while $\alpha<0$ accounts for under-dispersion. Note that in our use case, $\alpha$, like $\mu$ is estimated by the neural network and conditioned on the feature vector, allowing varying dispersion per-sample. Given the density function in Eq. \ref{eq:gen_poisson_dist}, the resultant log-likelihood loss for a dataset with $M$ samples is defined as:

\begin{align}
\label{eq:gen_poisson_loss}
    L_{\text{gp}}(X,Y) = - \frac{1}{M} \sum_{i=1}^M & \bigg[
            y^{(i)} \big(\log \mu^{(i)} - \log(1+\alpha^{(i)} \mu^{(i)})\big) \nonumber\\
             &+(y^{(i)} - 1)\log(1+\alpha^{(i)} y^{(i)}) \nonumber \\
             &- \frac{\mu^{(i)}(1 +\alpha^{(i)} y^{(i)})}
                   {1 + \alpha^{(i)} \mu^{(i)}} + + \log(y^{(i)}!)\bigg],
\end{align}
\noindent
where $\alpha^{(i)}$ and $\mu^{(i)}$ are obtained as transformed outputs of the neural network in a similar fashion as we obtain $\mu^{(i)}$ for the Poisson loss. In practice, as for the Poisson loss, we dropped the term related to $y!$ since it does not affect the optimization of the network parameters.

Note also that restrictions must be placed on the negative value of the $\alpha^{(i)}$ term to keep the arguments of the logarithm positive. For numerical convenience, we used an exponential activation over the dense layer for our $\alpha^{(i)}$ estimator, which accommodates over-dispersion but not under-dispersion.  Results from experiments comparing the use of Poisson and Generalized Poisson auxiliary losses are presented in Section \ref{sec:poisson_results}.

\subsection{Per-Vendor Malware Loss}
\label{sec:pervendormalwareloss}
The aggregation service from which we collected data for our experiments contains a breakout of individual vendor results per sample. We identified a subset $\mathcal{V} = \{v_1, \dots, v_V \}$ of 9 vendors that each produced a result for (nearly) every sample in our data. Each vendor result was added as a target in addition to the malware target by adding an extra fully connected layer per vendor followed by a sigmoid activation function to the end of the shared architecture. Standard binary cross-entropy loss per vendor is used for training. Note that this differs from the vendor count loss presented above in that each high-coverage vendor is used as an individual binary target, rather than being aggregated into a count.

The aggregate \emph{vendors} loss $L_{\text{vdr}}$ for the $V = 9$ selected vendors is simply the sum of the individual vendor losses:

\begin{align}
    L_{\text{vdr}}(X, Y) &= \frac{1}{M} \sum_{i=1}^{M} \sum_{j=1}^{V}
    \ell_{\text{vdr}}\left(f_{\text{vdr}_j} \big(\mathbf{x}^{(i)}\big),
                                  y_{v_j}^{(i)} \right)    \nonumber \\
     &= -\frac{1}{M} \sum_{i=1}^M \sum_{j=1}^{V}
     y_{v_j}^{(i)} \log(\hat{y}_{v_j}^{(i)}) +
     (1 - y_{v_j}^{(i)}) \log(1-\hat{y}_{v_j}^{(i)}),
\end{align}

\noindent
where $\ell_{\text{vdr}}$ is the per-sample binary cross-entropy function and $f_{\text{vdr}_j}\big(\mathbf{x}^{(i)}\big) = \hat{y}_{v_j}^{(i)}$ the output of the network that is trained to predict the label $y_{v_j}^{(i)}$ assigned by vendor $j$ to input sample $\mathbf{x}^{(i)}$.

Results from experiments exploring the use of individual vendor targets in addition to malware label targets are presented in Section \ref{sec:vendor_results}.

\subsection{Malicious Tags Loss}
\label{sec:tags_loss}
In this experiment we attempt exploit information contained in family detection names provided by different vendors in the form of malicious tags. We define a malicious tag as a high level description of the purpose of a given malicious sample. The tags used as auxiliary targets in our experiments are: \emph{flooder, downloader, dropper, ransomware, crypto-miner, worm, adware, spyware, packed, file-infector}, and \emph{installer}.

The process of creating these tags consists in parsing the individual vendor detection names to first extract relevant tokens within these detection names. We use a set of 10 vendors that we know, by experience, provide high quality detection names. Once we extracted the most common tokens from the detection names via simple parsing, we manually filtered them to keep only those tokens related to well-known malware family names or tokens that could be easily associated with one or more of our tags. For example, the token \emph{xmrig} -- even though it is not a malware family -- can be recognized as referring to a crypto-currency mining software and therefore can be associated with the \emph{crytpo-miner} tag. We then create a mapping from tokens to tags based on prior knowledge.
We label a sample as associated with tag $t_i$ if any of the tokens associated with $t_i$ are present in any of the detection names assigned to the sample by the set of trusted vendors.

%Once we annotate our dataset with these tags, we can define the tag prediction task as multi-label classification, 
Annotating our dataset with these tags, allows us to  define the tag prediction task as multi-label binary  classification,since zero or more tags from the set of possible tags  $\mathcal{T} = \{t_1,\dots, t_T \}$ can be present at the same time for a given sample. We introduce this prediction task in order to have targets in our loss function that are not not directly related to the number of vendors that recognize the sample as malicious. The vendor counts and the individual vendor labels are closely related with the definition of our main target, i.e. the malicious label, which classifies a sample as malicious if 5 or more vendors identify the sample as malware (see Section \ref{sec:mal_loss}). In the case of the tag targets, this information is not present. For instance, if all the vendors recognize a given sample as coming from the \emph{WannaCry} family in their detection names, the sample will be associated only once with the \emph{ransomware} tag. On the converse, because of our tagging mechanism, if only one vendor considers that a given sample is malicious and classifies it as coming from the \emph{WannaCry} family, the \emph{ransomware} tag will be present (although our malicious label will be 0). 

In order to predict these tags, we use a \emph{multi-headed} architecture in which we add two additional layers per tag to the end of the shared base architecture followed by a sigmoid activation function, as shown in Figure \ref{fig:schematic}. Each tag $t_j$ out of the possible $T=11$ tags has its own loss term computed with binary cross-entropy. Like the per-vendor malware loss, the aggregate tag loss is the sum of the individual tag losses. For the dataset with $M$ samples it becomes:

\begin{align}
    L_{\text{tag}}(X, Y) &=  \frac{1}{M} \sum_{i=1}^M \sum_{j=1}^{T}
    \ell_{\text{tag}}\left(
        f_{\text{tag}_j} \big(\mathbf{x}^{(i)}\big), y_{t_j}^{(i)}
    \right)
    \nonumber \\
     &= -\frac{1}{M} \sum_{i=1}^M \sum_{j=1}^{T}
     y_{t_j}^{(i)} \log(\hat{y}_{t_j}^{(i)}) +
                        (1 - y_{t_j}^{(i)}) \log(1-\hat{y}_{t_j}^{(i)}),
\end{align}
\noindent
where $y_{t_j}^{(i)}$ indicates if sample $i$ is annotated with tag $j$, and 
$\hat{y}_{t_j}^{(i)} = f_{\text{tag}_j} \big(\mathbf{x}^{(i)}\big)$
is the prediction issued by the network for that value.

 \subsection{Sample Weights}
\label{sec:sample_weights}

While our multi-objective network has the advantage that multiple labels and loss functions serve as additional sources of information, this introduces an additional complexity: given many (potentially missing) labels for each sample, we cannot rely on having all labels for a large quantity of the samples. Moreover, this problem gets worse as more labels are added. To address this, we incorporated per-sample weights, depending on the presence and absence of each label. For labels that are missing, we assign them to a default value and then set the associated weights to zero in the loss computation so a sample with a missing target label will not add to the loss computation for that target. Though this introduces slight implementation overhead, it allows us to train our network, even in the presence of partially labeled samples (e.g., when a vendor decides not to answer).

%buys us the added advantage of additional samples to train on, for example, when the aggregate malware labels do not exist (e.g., because 2-4 in our 1-/5+ model), we can still glean information from the auxiliary labels. We assess the effects of this \textit{partially supervised training} in Sec. \ref{sec:discussion}. 
%Given $M$ training samples and $K$ targets with respective per-sample training losses $l_k$, the loss over the dataset becomes

%\begin{equation}
%    L(X,Y) = \sum_{i=1}^{M} \sum_{k=1}^{K} w_{ik} l_k(\hat{y}_{ik},y_{ik})
%\end{equation}
%\noindent
%where $w_{ik}$ is simply an indicator function that is one if the label $k$ exists for sample $i$ ($y_{ik}$ is not missing) and zero otherwise.    

\subsection{Dataset}
\label{sec:dataset}

We collected two datasets of PE files and associated metadata from a threat intelligence feed: a set for training/validation and a test set. For the training/validation set, we pulled 20M PE files and associated metadata, randomly sub-selecting over a year -- from September 6, 2017 to September 6, 2018. For the test set, we pulled files from October 6, 2018 to December 6, 2018. Note also that we indexed files based on unique SHA for first seen time, so every PE in the test set comes temporally after the ones in the training set. We do not use a randomized cross-validation training/test split as is common in other fields, because that would allow the set on which the classifier was trained to contain files ``from the future'', leading to spuriously optimistic results. The reason for the one month gap between the end of the training/validation set and the start of the test set is to simulate a realistic worst-case deployment scenario where the detection model of interest is updated on a monthly basis. 

We then extracted $1024$-element  feature vectors for all those files using feature type described in \cite{saxe2015deep} and derived an aggregate malicious/benign label using a 1-/5+ criterion as described above. Invalid PE files were discarded. 

Of the valid PE files from which we were able to extract features we then further subsampled  our training dataset to 8,775,185 training samples, 
%9,000,000 training samples, 
with 7,188,150, or 81.9\% malicious and 1,587,035 or 18.1\% benign. %The remaining 224,815 or 2.5\% are \textit{gray} samples, without a benign or malicious label. These occur as a result of samples where the total number of vendor detections is between 2 and 4 and does not meet our 1-/5+ labeling criterion. 
Our validation set was also randomly subsampled from the same period as the training data and used to monitor convergence during training.  It consisted of 97,439 samples; of these, 17,620 were benign (18.1\%) and 79,819 were malicious (81.9\%).%, and 2,561 were gray (2.56\%).
Our test set exhibited similar statistics, with a size of %7,656,573 
7,368,851 total samples, 1,606,787 benign (21.8\%), and  5,762,064 malicious (78.2\%).%, and 287,722 gray (3.76\%).

%%%%%%%%%%%%%%%%%%%%%%%%%%%%%%%%%%%%%%%%%%%%%%
% 4. EXPERIMENTAL EVALUATION
%%%%%%%%%%%%%%%%%%%%%%%%%%%%%%%%%%%%%%%%%%%%%%

\section{Experimental Evaluation}
\label{sec:experiments}

In this section, we apply the auxiliary losses presented in in Section \ref{sec:implementation}, first individually, each loss in conjunction with a main malicious/benign loss, and then simultaneously in one combined model. We then compare to a baseline model, finding that each loss term yields an improvement, either in or in Receiver Operating Characteristic (ROC) net area under the curve (AUC) or in terms of detection performance at low false positive rates (FPR). We note that none of the auxiliary losses presented in this section, when added, reduced AUC or low-FPR ROC performance on the aggregate malicious/benign label from the baseline model. Each model used a loss weight of 1.0 on the aggregate malicious/benign loss and 0.1 on each auxiliary loss, i.e. when we add $K$ targets to the main loss, the final loss that gets backpropagated through the model becomes

\begin{align}
    L(X, Y) = L_{\text{mal}}(X, Y) + 0.1 \sum_{k=1}^K L_k(X, Y).
\end{align}

\begin{figure*}[!ht]
\begin{centering}
\subfloat[Count Loss]{\includegraphics[width=0.435\linewidth]{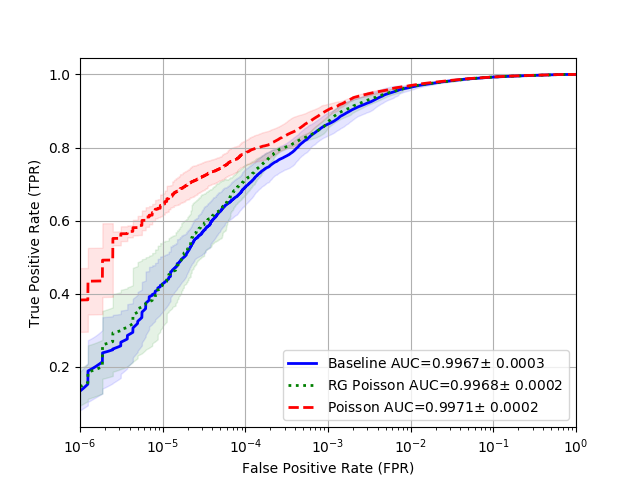}\label{fig:rgp_experiment}}
\subfloat[Vendor Loss]{\includegraphics[width=0.435\linewidth]{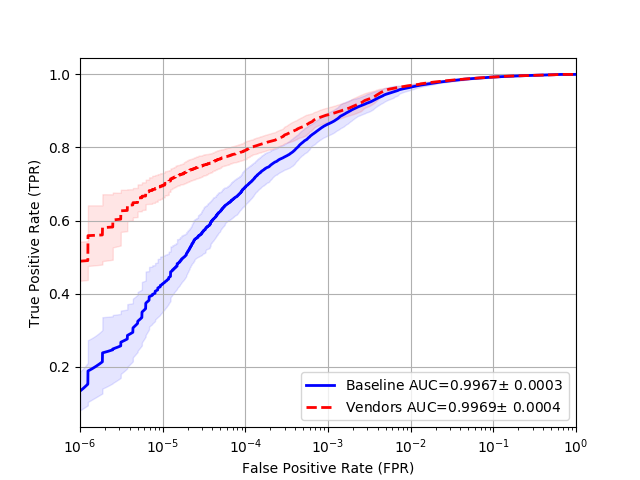}\label{fig:vendor_experiment}}\\
\subfloat[Tag Loss]{\includegraphics[width=0.435\linewidth]{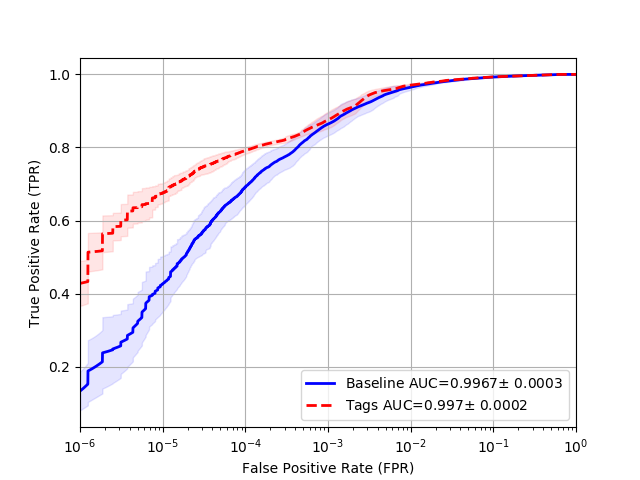}\label{fig:tag_experiment}}
\subfloat[Combined Loss]{\includegraphics[width=0.435\linewidth]{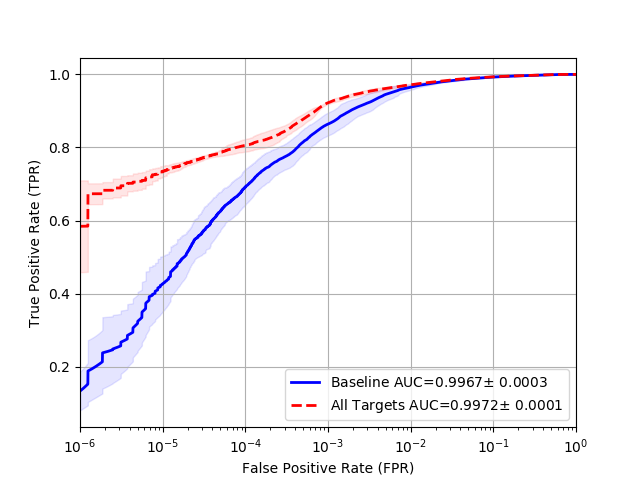}\label{fig:all_targets_roc}}
\caption[width=\linewidth]{ROC curves and AUC statistics for count, vendor, and tag losses compared to our baseline. Lines  represent the mean TPR at a given FPR, while shaded regions represent $\pm 1$ standard deviation. Statistics were aggregated over 5 training runs, each with random parameter initialization. \protect\subref{fig:rgp_experiment}  \emph{Count loss}. Our baseline model (blue solid line) is shown compared to a model employing a Poisson auxiliary loss (red dashed line), and a dispersed Poisson auxiliary loss (green dotted line). \protect\subref{fig:vendor_experiment} Auxiliary loss on multiple vendors malicious/benign labels (red dashed line) and baseline (blue solid line). \protect\subref{fig:tag_experiment} Auxiliary loss on semantic attribute tags (red dashed line) and baseline (blue solid line).
\protect\subref{fig:all_targets_roc} Our combined model (red dashed line) and baseline (blue solid line). The combined model utilizes an aggregate malicious/benign loss with an auxiliary Poisson count loss, a multi-vendor malicious/benign loss, and a malware attribute tag loss.
}
\label{fig:results}
\end{centering}
\end{figure*}

Results are depicted in graphical form in Figure \ref{fig:results} and in tabular form in Table \ref{tab:results}, with uncertainty evaluated by initializing each model's weights randomly and training for 10 epochs over five runs. Notice that the ROC curves in Figure \ref{fig:results} are plotted on a logarithmic scale for visibility, since the baseline performance is already quite high and significant marginal improvements are difficult to discern. For this reason, we also include relative reductions in mean true positive detection error and in standard deviation from the baseline for our best model in Table \ref{tab:results}, and for all models in Table \ref{tab:results_improvement_full} in the appendix.  

\begin{table*}[ht]
\centering
\begin{tabular}{llllll}

\cline{2-6}
           & \multicolumn{5}{|c|}{FPR}                               \\ 

\multicolumn{1}{l|}{}            & \multicolumn{1}{c|}{$10^{-5}$}               & \multicolumn{1}{c|}{$10^{-4}$}               & \multicolumn{1}{c|}{$10^{-3}$}         & \multicolumn{1}{c|}{$10^{-2}$}               & \multicolumn{1}{c|}{$10^{-1}$}               \\ \hline \hline

\multicolumn{1}{l||}{TPR Baseline}    & \multicolumn{1}{l|}{0.427 $\pm$ 0.076} & \multicolumn{1}{l|}{0.692 $\pm$ 0.049} & \multicolumn{1}{l|}{0.864 $\pm$ 0.031} & \multicolumn{1}{l|}{0.965 $\pm$ 0.007} & \multicolumn{1}{l|}{0.9928 $\pm$ 0.0007} \\ \hline

\multicolumn{1}{l||}{TPR Poisson}     & \multicolumn{1}{l|}{0.645 $\pm$ 0.029} & \multicolumn{1}{l|}{0.785 $\pm$ 0.034} & \multicolumn{1}{l|}{0.903 $\pm$ 0.016} & \multicolumn{1}{l|}{0.970 $\pm$ \textbf{0.001}} & \multicolumn{1}{l|}{0.9932 $\pm$ \textbf{0.0002}} \\ \hline

\multicolumn{1}{l||}{TPR RG Poisson}  & \multicolumn{1}{l|}{0.427 $\pm$ 0.116} & \multicolumn{1}{l|}{0.711 $\pm$ 0.041} & \multicolumn{1}{l|}{0.870 $\pm$ 0.016} & \multicolumn{1}{l|}{0.966 $\pm$ 0.003} & \multicolumn{1}{l|}{0.9930 $\pm$ 0.0003} \\ \hline

\multicolumn{1}{l||}{TPR Vendors}     & \multicolumn{1}{l|}{0.697 $\pm$ 0.034} & \multicolumn{1}{l|}{0.792 $\pm$ 0.024} & \multicolumn{1}{l|}{0.889 $\pm$ 0.020} & \multicolumn{1}{l|}{0.970 $\pm$ 0.004} & \multicolumn{1}{l|}{0.9928 $\pm$ 0.0014} \\ \hline

\multicolumn{1}{l||}{TPR Tags}        & \multicolumn{1}{l|}{0.677 $\pm$ 0.027} & \multicolumn{1}{l|}{0.792 $\pm$ \textbf{0.009}} & \multicolumn{1}{l|}{0.875 $\pm$ 0.022} & \multicolumn{1}{l|}{0.971 $\pm$ 0.004} & \multicolumn{1}{l|}{0.9932 $\pm$ 0.0008} \\ \hline

\multicolumn{1}{l||}{TPR All Targets} & \multicolumn{1}{l|}{\textbf{0.735} $\pm$ \textbf{0.014}} & \multicolumn{1}{l|}{\textbf{0.806} $\pm$ 0.017} & \multicolumn{1}{l|}{\textbf{0.922} $\pm$ \textbf{0.004}} & \multicolumn{1}{l|}{\textbf{0.972} $\pm$ 0.003} & \multicolumn{1}{l|}{\textbf{0.9934} $\pm$ 0.0004} \\ \hline \hline 
\multicolumn{1}{l||}{\% Error Reduction (All Targets)}    & \multicolumn{1}{c|}{\textbf{53.8\%}} & \multicolumn{1}{c|}{\textbf{37.0\%}} & \multicolumn{1}{c|}{\textbf{42.7\%} } & \multicolumn{1}{c|}{\textbf{20.0\%}} & \multicolumn{1}{c|}{\textbf{\ 8.3\%}} \\ \hline

\multicolumn{1}{l||}{\% Variance Reduction (All Targets)}    & \multicolumn{1}{c|}{\textbf{81.6\%}} &
\multicolumn{1}{c|}{65.3\%} &
\multicolumn{1}{c|}{\textbf{87.1\%}} &
\multicolumn{1}{c|}{57.1\%} & 
\multicolumn{1}{c|}{94.3\%} \\ \hline
\end{tabular}
\caption{Top: Mean and standard deviation true positive rates (TPRs) for the different experiments in Section  \ref{sec:experiments} at false positive rates (FPRs) of interest. Results were aggregated over five training runs with different weight initializations and minibatch orderings. Best results consistently occurred when using all auxiliary losses and are shown in bold. Bottom: Percentage reduction in missed true positive detections and percentage reductions in ROC curve standard deviation resulting from the best model (All Targets) compared to the baseline across various FPRs. State-of-the-art results are shown in \textbf{bold}.}\label{tab:results}
\end{table*}

%\begin{figure*}
%\begin{subfigure}{0.31\textwidth}
%\includegraphics[width=\linewidth]{images/rgp_with_uncertainty.png}
%\caption{First subfigure} \label{fig:1a}
%%\end{subfigure}
%\hspace*{\fill} % separation between the subfigures
%\begin{subfigure}{0.31\textwidth}
%\includegraphics[width=\linewidth]{images/tags_expts.png}
%\caption{Second subfigure} \label{fig:1b}
%\end{subfigure}
%\hspace*{\fill} % separation between the subfigures
%\begin{subfigure}{0.31\textwidth}
%\includegraphics[width=\linewidth]{images/tags_expts.png}
%\caption{Third subfigure} \label{fig:1c}
%\end{subfigure}
%\caption{A figure that contains three subfigures} \label{fig:1}
%\end{figure*}

\subsection{Vendor Count Loss}
\label{sec:poisson_results}

We employed the same base PE model topology as for our other experiments, with a primary malicious/benign binary cross entropy loss, and an auxiliary count loss. We experimented with two different loss functions for the count loss -- a Poisson loss and a Restricted Generalized Poisson loss (equations \ref{eq:poisson_loss} and \ref{eq:gen_poisson_loss} respectively). For the Poisson loss, we used an exponential activation over a dense layer atop the base to estimate $\mu^{(i)}$. For the Restricted Generalized Poisson loss, we followed a similar pattern using two separate dense layers with exponential activations on top; one for the $\mu^{(i)}$ parameter and another for the $\alpha^{(i)}$ parameter. The choice of an exponential activation is consistent with statistics literature on Generalized Linear Models (GLMs) \cite{mccullagh2018generalized}. 

Results on the malware detection task, using Poisson and Restricted Generalized Poisson (RG-Poisson) losses as an auxiliary loss function are shown in Figure \ref{fig:rgp_experiment}. When compared to a baseline using no auxiliary loss, we see a statistically significant improvement with the Poisson loss function in both AUC and ROC curve, particularly in low false positive rate (FPR) regions. The RG-Poisson loss, by contrast yields no statistically significant gains over the baseline in terms of AUC, nor does it appear to yield statistically significant gains at any point along the ROC curve. 

This suggests that the RG-Poisson loss model is ill-fit, which could stem from a variety of issues. First,  
%if dispersion issues, where they exist, are a consequence of under-dispersion, not over-dispersion -- this could occur because counts are bounded by the net number of vendors, and if vendors disproportionately trigger simultaneously; 
if counts are under-dispersed, an over-dispersed Poisson loss could be an inappropriate model. Under-dispersion could occur if certain vendors disproportionately trigger simultaneously or because counts are inherently bounded by the net number of vendors. Second, a Poisson model, even with added dispersion parameters, is an ill-posed model of count data, but removing the dispersion parameter removes a dimension in the parameter space to over-fit on. Inspecting the dispersion parameters predicted by the RG-Poisson model, we noted that they were relatively large, which supports this hypothesis. We also noticed that the RG-Poisson model converged significantly faster than the Poisson model in terms of malware detection loss. 

\subsection{Modeling Individual Vendor Responses}
\label{sec:vendor_results}

Incorporating the multi-label binary cross-entropy loss across vendors as an auxiliary loss, described in section \ref{sec:pervendormalwareloss}, in conjunction with the main malicious/benign loss yields a similar increase in the TPR at low FPR regions of the ROC curve (see Figure \ref{fig:vendor_experiment}) as the Poisson experiment. While this does not lead to a significant increase in AUC due to the fact that the improvement is integrated across an extremely narrow range of FPRs, this improvement in TPR at lower FPRs may still be operationally significant, and does indicate an improvement in the model.

\subsection{Incorporating Tags as Targets}

In this experiment we extend the architecture of our base network to predict, not only the malware/benign label, but also the set of 11 tags defined in section \ref{sec:tags_loss}.  For this, we add two fully connected layers (FC[512-256] and FC[256-1]) per tag to the end of the base architecture that will be trained to identify that tag from the shared representation. Each of these tag-specialized layers predicts a binary output for the presence or absence of the tag given the shared representation and uses a binary cross-entropy loss that gets added to the losses of the rest of the tags and the malicious/benign loss, i.e. the loss for this experiment is a sum containing one term per tag, weighted by a loss weight of 0.1 as mentioned at the beginning of this section, and one term for the loss incurred on the main task.

The result of this experiment represented via the ROC curves of Figure \ref{fig:tag_experiment}. Similar to section \ref{sec:vendor_results} we see no statistical difference in the AUC values with respect to the baseline, but we do observe substantial statistical improvement in the predictions of the model in low FPR regions, particularly for FPR values lower than $10^{-3}$. Furthermore, we also witness a substantial decrease in the variance of the ROC curve.

\subsection{Combined Model}
\label{sec:combined_model_results}
Finally, we extend our model to predict all auxiliary targets in conjunction with the aggregate label, with a net loss term containing a combination of all auxiliary losses used in previous experiments. The final loss function for the experiment is the sum of all the individual losses where the malware/benign loss has a weight of 1.0 while the rest of the losses have a weight of 0.1.

The resulting ROC curve and AUC are shown in Figure \ref{fig:all_targets_roc}. The AUC of $0.9972 \pm 0.0001$ is the highest obtained across all the experiments conducted in this study. Moreover, in contrast to utilizing any single auxiliary loss, we see a noticeable improvement in the ROC curve not only in very low FPR regions, but also at $10^{-3}$ FPR. Additionally, variance is consistently lower across a range of low-FPR values for this combined model than for our baseline or any previous models. An exception is  near $10^{-6}$ FPR where measuring variance is an ill-posed problem because even with a test dataset of over 7M samples, detecting or misdetecting even one or two of them can significantly affect detection rate. 

\section{Discussion}
\label{sec:discussion}
\vspace{-1em}
In this section, we examine the effects of  different types of auxiliary loss functions on main task ROC curve. We then perform a sanity check to establish whether our performance increases result from additional information added to our neural networks by the auxiliary loss functions or are the artifact of some regularization effect. 

\subsection{Modes of Improvement}
\label{sec:improvment}

Examining the plots in Figure \ref{fig:results}, we see three different types of improvement that result from our auxiliary losses:

\begin{enumerate}
\item A bump in TPR at low FPR ($< 10^{-3}$).\label{imp:tpr}
\item A net increase in AUC and a small bump in performance at higher FPRs ($\geq 10^{-3}$).\label{imp:auc}
\item A reduction in variance.\label{imp:var}
\end{enumerate}

Improvement \ref{imp:tpr} is particularly pronounced in the plots due to the logarithmic scale, but it does not substantially contribute to net AUC due to the narrow FPR range. However, this low-FPR part of the ROC is important from an operational perspective when deploying a malware detection model in practice. Substantially higher TPRs at low FPR could allow for novel use cases in an operational scenario. Notice that this effect is more pronounced for auxiliary losses containing multi-objective binary labels (Figs. \ref{fig:vendor_experiment}, \ref{fig:tag_experiment}, and \ref{fig:all_targets_roc}) than for the Poisson loss, suggesting that it occurs most prominently when employing our multi-objective binary label losses. 
Let us consider why a multi-objective binary loss might cause such an effect to occur: At low FPRs, we see high thresholds on the detection score from the main output of the network. To surpass this threshold and register as a detection, the main sigmoid output must be very close to 1.0, i.e., very high response. 
Under a latent correlation with the main output, a high-response hit for an auxiliary target label could also boost the response for the main detector output, while a baseline model without this information might mis-detect. We hypothesize that this improvement \ref{imp:tpr} occurs from having many objectives simultaneously and thereby increasing chances for a high-response target hit. The loss type may or may not be incidental, which is consistent with its noticeable but less pronounced presence under a single-objective Poisson auxiliary loss (Figure \ref{fig:rgp_experiment}).

Improvement \ref{imp:auc} likely stems from improvements in detection rate that we see around $10^{-3}$ FPR and higher. Notice that these effects are more pronounced in Figs. \ref{fig:rgp_experiment} and  \ref{fig:all_targets_roc}, are somewhat noticeable in Figure  \ref{fig:vendor_experiment}, and are not noticeable in Figure \ref{fig:tag_experiment}, consistent with the resultant AUCs. This  suggests that the effect occurs most prominently in the presence of an auxiliary count loss. We postulate that this occurs because our aggregate detection label is derived by thresholding the net number of vendor detections for each sample but in doing so removes a notional view of confidence that a sample is malicious, or alternatively difficulty of classifying a malicious sample or extent of ``maliciousness'' that the number of detection counts provides. Bear in mind that some detectors are better at detecting different types of malware than others, so more detections suggest a \textit{more malicious} file, e.g., with more malware components, or more widely blacklisted file (higher confidence). Providing information on the number of counts in an auxiliary loss function may therefore provide the classifier more principled information on how to order detection scores, thus allowing for more effective thresholding and a better ROC curve.

\begin{figure}[!t]
    \centering
    \includegraphics[width=\linewidth]{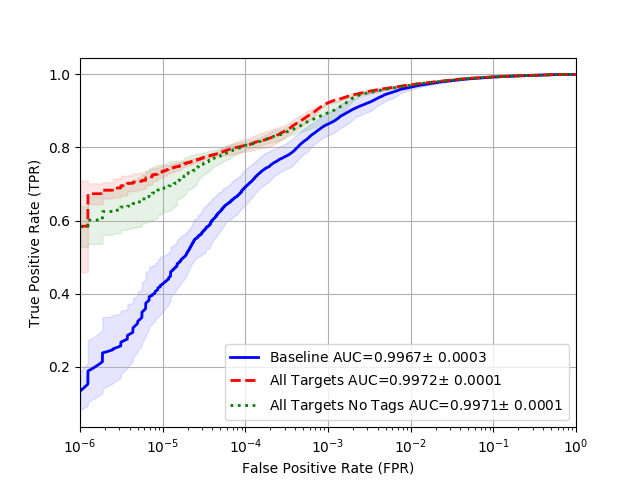}
    \caption{When we remove the attribute tags loss (green dotted line) we get a similar shaped ROC curve with similar ROC compared to using all losses (red dashed line), but with slightly higher variance in the ROC. This supports our hypotheses about effects of different loss functions on the shape of the ROC curve. The baseline is shown as a blue solid line for comparison.}\label{fig:ablation}
\end{figure}

Improvement \ref{imp:var} occurs across all loss types, particularly in low FPR ranges, with the exception of \textit{very low} FPRs (e.g., $10^{-6}$), where accurately measuring mean and variance is an ill-posed problem due to the size of the dataset (cf. section \ref{sec:combined_model_results}). Comparing the ROC plots in Figure \ref{fig:results}, the reduction in variance appears more pronounced as the number of losses increases. Intuitively, this is not a surprising result since adding objectives/tasks imposes constrains the allowable weight space -- while many choices of weights might allow a network to perform a single task well only a subset of these choices will work well for all tasks simultaneously. Thus, assuming equivalent base topology, we expect a network that is able to perform at least as well on multiple tasks as many single-task networks to exhibit lower variance.

Combining all losses seems to accentuates all improvements (\ref{imp:tpr}-\ref{imp:var}) with predictable modes which we attribute to our various loss types (Figure \ref{fig:all_targets_roc}) -- higher detection rate at low FPR brought about primarily by multi-objective binary losses,  a net AUC increase and a detection bump at $10^{-3}$ FPR brought about by the count loss, and a reduced variance brought about by many loss functions. To convince ourselves that this is not a coincidence, we also trained a network using only Poisson and vendor auxiliary losses but no attribute tags (cf Figure \ref{fig:ablation}). As expected, we see that this curve exhibits similar general shape and AUC characteristics that occur when training with all loss terms, but the variance appears slightly increased.  

In the variance-reduction sense, we can view our auxiliary losses as regularizers. This raises an important question: are improvements \ref{imp:tpr} and \ref{imp:auc} actually occurring for the reasons that we hypothesize or are they merely naive result of regularization? 

\subsection{Representation or Regularization?}

While the introduction of some kinds of auxiliary targets appears to improve the performance of the model on the main task, it is less clear why this is the case.  The reduction in variance produced by the addition of extra targets suggests one potential alternative explanation for the observed improvement: rather than inducing a more discriminative representation in the hidden layers of the network, the additional targets may be acting as constraints on the network, reducing the space of viable weights for the final trained network, and thus acting as a form of additional regularization.  Alternatively, the addition of extra targets may simply be accelerating training by amplifying the gradient; while this seems unlikely given our use of a validation set to monitor approximate convergence, we nevertheless also investigate this possibility.

To evaluate these hypotheses, we constructed three additional targets (and associated loss functions) that provided uninformative targets to the model: i) a pseudo-random target that is approximately independent of either the input features or the malware/benign label; ii) an additional copy of the main malware target transformed to act as a regression target; and iii) an additional copy of the main malware target.

The random target approach attempts to directly evaluate whether or not an additional pseudo-random target might improve network performance by `using up' excess capacity that might otherwise lead to overfitting.  We generate pseudo-random labels for each sample based off of the parity of a hash of the file contents. While this value is effectively random and independent of the actual malware/benignware label of the file, the use of a hash value ensures that a given sample will always produce the same pseudo-random target.  This target is fit via standard binary cross-entropy loss against a sigmoid output,

\begin{align}
    L_{\text{rnd}}(X, Y) =
        -\frac{1}{M} \sum_{i=1}^M &
            y^{(i)} \log f_{\text{rnd}}(\mathbf{x}^{(i)}) + \nonumber \\
            &(1- y^{(i)}) \log \left(1- f_{\text{rnd}}(\mathbf{x}^{(i)})\right),
\end{align}
\noindent
where $f_{\text{rnd}}\left( \mathbf{x}^{(i)} \right)$ is the output of the network which is being fit to the random target $y^{(i)}$.

In contrast, the duplicated regression target evaluates whether further constraining the weights \emph{without} requiring excess capacity to model additional independent targets has an effect on the performance on the main task. The model is forced to adopt an internal representation that can satisfy two different loss functions for perfectly correlated targets, thus inducing a constraint that does not add additional information.  To do this, we convert our binary labels (taking on values of 0 and 1 for benign and malware, respectively) to -10 and 10, and add them as additional \emph{regression} targets fit via mean squared error (MSE). Taking $y^{(i)}$ as the $i^{th}$ binary target and $f_{\text{MSE}}(\mathbf{x}^{(i)})$ as the regression output of the network, we can express the MSE loss as:

\begin{equation}
    L_{\text{mse}}(X, Y) =
    \sum_{i=1}^M
    \left(f_{\text{mse}}
    \left(\mathbf{x}^{(i)}\right) - 20\left(y^{(i)} - 0.5\right)\right)^2.
\end{equation}
\noindent

Finally, in the case of the duplicated target, the model effectively receives a larger gradient due to our duplication of the loss. The loss for the duplicated label uses the identical cross-entropy loss as for the main target, obtained by substituting $f_{\text{dup}}(\mathbf{x}^{(i)})$ for $f_{\text{mal}}(\mathbf{x}^{(i)})$ in equation \ref{eq:main_mal_loss} as the additional model output that is being fit to the duplicated target.
%\endnote{
Note that we performed two variants on the duplicated target experiment: one in which both the dense layer prior to the main malware target and the dense layer prior to the duplicated target were trainable, and one in which the dense layer for the duplicate target was frozen at its initialization values to avoid the trivial solution in which the pre-activation layer for both the main and duplicate target were identical.  In both cases, the results were equivalent; only results for the trainable case are shown.%}

Both $f_{\text{rnd}}$ and $f_{\text{dup}}$ are obtained by applying a dense layer followed by a sigmoid activation function to the intermediate output of the input sample from the shared base layer ($\mathbf{h}$ in Figure \ref{fig:schematic}), while $f_{\text{mse}}(\mathbf{x}^{(i)})$ is obtained by passing the intermediate representation of the input sample $\mathbf{h}$ through a fully connected layer with no output non-linearity.

\begin{figure}[!ht]
    \centering
    \includegraphics[width=\linewidth]{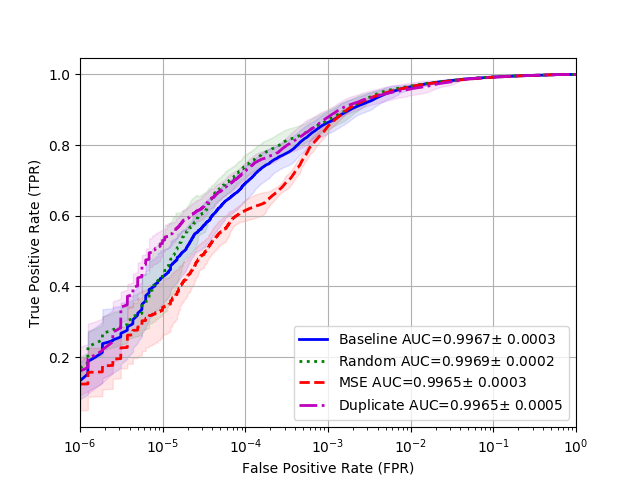}
    \caption{ROC curves comparing classification capabilities of models on the malware target when either random (green dotted line), regression (red dashed line), or duplicated targets (magenta dashed and dotted line) are added as auxiliary losses. Note that with the exception of the regression loss -- which appears to harm performance -- there is little discernible difference between the remaining ROC curves. The baseline is shown as a blue solid line for comparison.}
    \label{fig:noise_experiment}
\end{figure}

Results of all three experiments are shown in Figure \ref{fig:noise_experiment}. In no case did the performance of the model on the main task improve statistically significantly over the baseline. This suggests that auxiliary tasks must encode relevant information to improve the model's performance on the main task. For each of the three auxiliary loss types in Figure \ref{fig:noise_experiment}, there is no additional information provided by the auxiliary targets: the random target is completely uncorrelated from any information in the file (and indeed the final layer is ultimately dominated by the bias weights and produces a constant output of 0.5 regardless of the inputs to the layer), while the duplicated and MSE layers are perfectly correlated with the final target.  In either case, there is no incentive for the network to develop a richer representation in layers closer to the input; the final layer alone is sufficient given an adequate representation in the core of the model.

\section{Related Work}
\label{sec:relatedwork}
The application of machine learning to computer security dates back to the 1990's \cite{rudd2017survey}, but large-scale commercial deployments of deep neural networks (DNNs) that have led to transformative performance increases are a more recent phenomenon. Several works from the ML-Sec community have leveraged DNNs for statically detecting malicious content across a variety of different formats and file types \cite{saxe2015deep,saxe2018deep,rudd2018meade}. However, these works predominantly focus on applying regularized cross entropy loss functions between single network outputs and malicious/benign labels per-sample, leaving the potential of multiple-objective optimization largely untapped. A notable exception is \cite{huang2016mtnet}, where Huang et al. added a multiclass label for Microsoft's malware families, and used a categorical cross entropy loss function atop a softmax output as an auxiliary objective. Note that this is a heterogeneous classification task, not multiple sources of the same label, unlike our vendor label loss, and unlike our malware attribute tag loss, their labels are mutually exclusive. They also do not address the problem of missing labels. 

Despite the lack of attention from the ML-Sec community, multi-objective/multi-task optimization has been applied to other areas of machine learning for a long time, even if not explicitly referred to as such. The work of Abu-Mostafa \cite{abu1990learning} predates most explicit references to multi-task learning by introducing the concept of \emph{hints}, in which known invariances of a solution (such as traslation invariance, or invariance under negation) can be incorporated into network structure and used to generate additional training samples by applying the invariant operation to the existing samples, or -- most relevant to our work -- used as an additional target by enforcing that samples modified by an invariant function should not only be both correctly classified, but also explicitly classified identically. Caruna~\cite{caruna1993multitask} first introduced multi-task learning in neural networks as a ``source of inductive bias'' (also reframed as inductive transfer in \cite{caruana1998dozen}), in which more difficult tasks could be combined in order to exploit similarities between tasks that could serve as complementary signals during training.  While his work predates the general availability of modern GPUs, and thus the models and tasks he examines are fairly simple, Caruna nevertheless demonstrates that jointly learning related tasks produce better generalization on a task-by-task basis than learning them individually. It is interesting to note that in \cite{caruna1993multitask} he also demonstrated that learning multiple copies of the same task can also lead to a modest improvement in performance (which we did not observe in this work, possibly due to the larger scale and complexity of our task). 

Kumar and Duame \cite{kumar2012learning} consider a refinement on the basic multi-task learning approach that leads to clustering related tasks, in an effort to mitigate the potential of \emph{negative transfer} in which adding an unrelated task to the set of tasks being trained into a given model can degrade performance on the target task.  Similarly, the work of Rudd et al. \cite{rudd2016moon} explores the use of domain-adaptive weighting of tasks during the training process.

Multi-objective optimization has been applied to extremely complex image classification tasks, including predicting characters and ngrams within unconstrained images of text \cite{jaderberg2014deep}, joint facial landmark localization and detection \cite{ranjan2017hyperface}, image tagging and retrieval \cite{huang2015unconstrained, wu2015weakly}, and attribute prediction \cite{abdulnabi2015multi, rudd2016moon} where a common auxiliary task is to challenge the network to classify additional attributes of the image, such as manner of dress for full-body images of people or facial attributes (e.g., smiling, narrow eyes, race, gender).  While a range of possible network structures are possible, common exemplars are having largely independent networks with a limited number of shared weights (as in \cite{abdulnabi2015multi}), a single network with minimal separation between tasks (as in \cite{rudd2016moon}), or a number of parallel single-task classifiers in which the weights are constrained to be similar to each other. A more complex approach may be found in \cite{ruder122017sluice}, in which the sharing between tasks is learned automatically in an on-line fashion.

Other, more distantly connected  domains of machine learning research reinforce the intuition that learning on disparate tasks can improve model performance. Work in semi-supervised learning, such as \cite{kingma2014semi} and \cite{rasmus2015semi}, has shown the value of additional reconstruction and denoising tasks to learn representations valuable for a core classification model, both through regularization and through access to a larger dataset than is available with label. The widespread success of transfer learning is also a testament to the value of training a single model on nominally distinct tasks. BERT \cite{devlin2018bert}, a recent example from the Natural Language Processing literature, shows strong performance gains from pre-training a model on masked-word prediction and predictions of whether two sentences appear in sequence, even when the true task of interest is quite distinct (e.g. question answering, translation). 
 
Multi-view learning (see \cite{xu2013survey} for a survey) is a related approach in which multiple inputs are trained to a single target. This approach also arguably leads to the same general mechanism for improvement: the model is forced to learn relationships between sets of features that improve the performance using any particular set.  While this approach often requires all sets of features to be available at test time, there are other approaches, such as \cite{vapnik2015learning}, that relax this constraint.

\section{Conclusion}
\label{sec:conclusion}
In this paper, we have demonstrated the effectiveness of auxiliary losses for malware classification. We have also provided experimental evidence which suggests that performance gains result from an improved and more informative representation, not merely a regularization artifact. This is consistent with our observation that improvements occur as additional auxiliary losses and different loss types are added. We also note that different loss types have different effects on the ROC; multi-label vendor and semantic attribute tag losses have greatest effect at low false positive rates ($\leq 10^{-3}$), while Poisson counts have a substantial net impact on AUC, the bulk of which stems from detection boosts at higher FPR. 

While we experimented on PE malware in this paper, our auxiliary loss technique could be applied to many other problems in the ML-Sec community, including utilizing a label on format/file type for format-agnostic features (e.g., office document type in \cite{rudd2018meade}) or file type under a given format, for example APKs and JARs both share an underlying ZIP format; a zip-archive malware detector could use tags on the file type for auxiliary targets. Additionally, tags on topics and classifications of embedded URLs could serve as auxiliary targets when classifying emails or websites.

One open question is whether or not multiple auxiliary losses improve each others' performances as well as the main target. If the multiple outputs of operational interest (such as the tagging output) can be trained simultaneously while also increasing (or at least not decreasing) their joint accuracy, this could lead to models that are both more compact and more accurate than individually deployed ones. In addition to potential accuracy gains, this has significant potential operational benefits, particularly when it comes to model deployment and updates. We defer a more complete evaluation of this question to future work.

%We have not evaluated how the auxiliary impact each other with respect to accuracy, largely because predictions corresponding to the associated losses have little immediate application, with exception of attribute tags, which could be used, for example, in and endpoint detection and response (EDR) context. We leave this analysis for future research, but hypothesize that most targets benefit from auxiliary loss functions assuming that the auxiliary targets are informative. If this is the case, this is a strong argument for multi-target deployed models -- e.g., a combined \textit{detection and tagging} model -- because the combined model is more compact and more accurate than multiple separate models.

While this work has focused on applying auxiliary losses in the context of deep neural networks, there is nothing mathematically that precludes using them in conjunction with a number of other classifier types. Notably, gradient boosted classifier ensembles, which are also popular in the ML-Sec community could take very similar auxiliary loss functions even though the structure of these classifiers is much different. We encourage the ML-Sec research community to implement multi-objective ensemble classifiers and compare with our results. Our choice of deep neural networks for this paper is infrastructural more than anything else; while several deep learning platforms, including PyTorch, Keras, and Tensorflow among others easily support multiple objectives and custom loss functions, popular boosting frameworks such as lightGBM and XGBoost have yet to implement this functionality.  

The analyses conducted herein used metadata that can naturally be transformed into a label source and impart additional information to the classifier with no extra data collection burden on behalf of the threat intelligence feed. Moreover, our auxiliary loss technique does not change the underlying feature space representation. 
Other types of metadata, e.g., the file path of the malicious binary or URLs extracted from within the binary might be more useful in a multi-view context, serving as input to the classifier, but this approach raises challenges associated with missing data that our loss weighting scheme trivially addresses. Perhaps our weighting scheme could even be extended, e.g., by weighting each sample's loss contribution according to certainty/uncertainty in that sample's label, or re-balancing the per-task loss according to the expected frequency of the label in the target distribution. This could open up novel applications, e.g., detectors customized to a particular user endpoints and remove sampling biases inherent to multi-task data.
 
%The degree to which detectors trained with auxiliary loss terms are robust to concept drift/time decay is another evaluation which we leave to future research. In our experiments, we introduced a one month time lag between training and test samples, which we believe reasonably reflects operational assumptions for most vendors -- that they will re-deploy a freshly trained model every month or so. This suggests that our auxiliary loss model is ``robust enough" to be useful in operational settings. However, the time decay characteristics of a multi-target vs. a single-target model could have ramifications on a principled choice of re-deployment time. 

%\section{Acknowledgments}
%This work was sponsored by Sophos PLC.
%\td{Leave this out for the review draft.}

%\section{Availability}

%It's great when this section says that MyWonderfulApp is free software, 
%available via anonymous FTP from

%%\begin{center}
%{\tt ftp.site.dom/pub/myname/Wonderful}\\
%\end{center}

%Also, it's even greater when you can write that information is also 
%available on the Wonderful homepage at 

%\begin{center}
%{\tt http://www.site.dom/\~{}myname/SWIG}
%\end{center}

{\normalsize \bibliographystyle{acm}
\bibliography{references}}

%\theendnotes
\counterwithin{table}{section}
\newpage
\onecolumn
\appendix
\section{Appendix}

\begin{table*}[ht]
\centering

\begin{tabular}{llllll}
\cline{2-6}
           & \multicolumn{5}{|c|}{FPR}                               \\ 

\multicolumn{1}{l|}{}            & \multicolumn{1}{c|}{$10^{-5}$}               & \multicolumn{1}{c|}{$10^{-4}$}               & \multicolumn{1}{c|}{$10^{-3}$}         & \multicolumn{1}{c|}{$10^{-2}$}               & \multicolumn{1}{c|}{$10^{-1}$}               \\ \hline \hline

\multicolumn{1}{l||}{Poisson}    & \multicolumn{1}{l|}{38.05, 61.84} & \multicolumn{1}{l|}{30.19, 30.61} & \multicolumn{1}{l|}{28.68, 48.39} & \multicolumn{1}{l|}{14.29,  \textbf{85.71}} & \multicolumn{1}{l|}{\ 5.56, \textbf{97.14}} \\ \hline

\multicolumn{1}{l||}{RG Poisson}    & \multicolumn{1}{l|}{\ 0.00, -52.63} & \multicolumn{1}{l|}{\ 6.17, 16.33} & \multicolumn{1}{l|}{\ 4.41, 48.39} & \multicolumn{1}{l|}{\ 2.86,  57.14} & \multicolumn{1}{l|}{\ 2.78, 95.71} \\ \hline

\multicolumn{1}{l||}{Vendors}    & \multicolumn{1}{l|}{47.12, 55.26} & \multicolumn{1}{l|}{32.47, 51.02} & \multicolumn{1}{l|}{18.38, 35.48} & \multicolumn{1}{l|}{14.29,  42.86} & \multicolumn{1}{l|}{\ 0.00, 80.00} \\ \hline

\multicolumn{1}{l||}{Tags}    & \multicolumn{1}{l|}{43.63, 64.47} & \multicolumn{1}{l|}{32.47, \textbf{81.63}} & \multicolumn{1}{l|}{\ 8.09, 29.03} & \multicolumn{1}{l|}{17.14, 42.86} & \multicolumn{1}{l|}{\ 5.56, 88.57} \\ \hline

\multicolumn{1}{l||}{All Targets}    & \multicolumn{1}{l|}{\textbf{53.75}, \textbf{81.58}} & \multicolumn{1}{l|}{\textbf{37.01}, 65.31} & \multicolumn{1}{l|}{\textbf{42.65}, \textbf{87.10}} & \multicolumn{1}{l|}{\textbf{20.00}, 57.14} & \multicolumn{1}{l|}{\textbf{\ 8.33}, 94.29} \\ \hline
\end{tabular}
\caption{Relative percentage reductions in true positive detection error and standard deviation compared to the baseline model (displayed as \textit{detection error reduction}, \textit{standard deviation reduction}) at different false positive rates (FPRs) for the different experiments in section \ref{sec:experiments}. Results were evaluated over five different weight initializations and minibatch orderings. Best detection error reduction consistently occurred when using all auxiliary losses. Best results are shown in \textbf{bold}.}\label{tab:results_improvement_full}
\end{table*}

\end{document}